\documentclass[prc,twocolumn,showpacs,showkeys,nofootinbib,superscriptaddress]{revtex4-1}
\usepackage{graphicx}
\usepackage{dcolumn}
\usepackage{epsfig}
\usepackage{bm}

\newcommand{\eq}{\begin{eqnarray}}
\newcommand{\en}{\end{eqnarray}}

\newcommand{\ba}[1]{\begin{eqnarray} \label{(#1)}}
\newcommand{\ea}{\end{eqnarray}}

\begin{document}

\topmargin -0.50in

\title{Few active mechanisms of the $0\nu\beta\beta$-decay and 
effective mass of Majorana neutrinos}

\author{Fedor \v Simkovic}
\affiliation{Laboratory of Theoretical Physics, JINR,
141980 Dubna, Moscow region, Russia}
\affiliation{Department of Nuclear Physics and Biophysics, 
Comenius University, Mlynsk\'a dolina F1, SK--842 15
Bratislava, Slovakia}
\author{John Vergados}
\affiliation{Theoretical Physics Division, University of Ioannina, 
GR--451 10 Ioannina, Greece}
\author{Amand Faessler}
\affiliation{Institute f\"{u}r Theoretische Physik der Universit\"{a}t
T\"{u}bingen, D-72076 T\"{u}bingen, Germany}

\begin{abstract}
It is well known that there exist many mechanisms that may contribute to neutrinoless double beta decay. By exploiting the fact that the associated nuclear matrix elements are target dependent we show that, given definite experimental results on a sufficient number of targets, one can determine or sufficiently constrain all  lepton violating parameters including the mass term. As a specific example we show that, assuming the observation of the $0\nu\beta\beta$-decay 
in three different nuclei, e.g. $^{76}$Ge, $^{100}$Mo and $^{130}$Te, and just
 three lepton number violating mechanisms (light and heavy
neutrino mass mechanisms as well as R-parity breaking SUSY mechanism) being active, there are only four different solutions for the lepton violating parameters, provided that they are relatively real. In particular, assuming evidence of the $0\nu\beta\beta$-decay
of $^{76}$Ge, the effective neutrino  Majorana mass
$|m_{\beta\beta}|$ can be almost uniquely extracted  by utilizing  other existing constraints (cosmological
observations and tritium $\beta$-decay experiments).
We also point out the possibility that the non-observation
of the $0\nu\beta\beta$-decay for some isotopes could be 
in agreement with a value of $|m_{\beta\beta}|$ in sub eV region. We thus suggest that
it is important to have at least two different $0\nu\beta\beta$-decay 
experiments for a given nucleus.
\end{abstract}



\date{\today}

\maketitle

\section{Introduction}

After the discoveries of oscillations of  atmospheric, solar and
terrestrial neutrinos, one has gained a lot of valuable information regarding the mixing matrix and the squared mass differences. The absolute scale of the neutrino mass cannot, however, be determined in such experiments. Our best hope for settling this important issue as well as solving a second
 challenging problem, i.e. whether the neutrinos are Majorana or Dirac particles, is the observation of neutrinoless double beta decay.

The total lepton number violating 
neutrinoless double beta decay ($0\nu\beta\beta$-decay),
\begin{equation}
(A,Z) \rightarrow (A,Z+2) + 2 e^-,
\end{equation}
can take place only if the neutrino  
is a massive Majorana particle \cite{schv}. The measurement of
the $0\nu\beta\beta$-decay rate could, in principle, determine an absolute 
scale of neutrino mass, solve the neutrino mass hierarchy problem and provide 
information about the Majorana CP-violating phases of neutrinos.

The evidence for a $0\nu\beta\beta$-decay of ${^{76}Ge}$ 
has been claimed by some authors of the Heidelberg-Moscow 
collaboration at Laboratori Nazionali del Gran Sasso
\cite{evidence} with 
\begin{equation}
T^{0\nu}_{1/2}(^{76}Ge)=2.23^{+0.44}_{-0.31}\times 10^{25}~\mbox{y}. 
\label{evid}
\end{equation}
Such a claim has raised some criticism but none of the existing
experiments can rule it out \cite{elisi}. The only certain way 
to confirm or refute this claim is with additional sensitive 
experiments \cite{avig}, in particular the GERDA experiment 
\cite{gerda}, which plans to start taking data this year. 

There is a general consensus that the $0\nu\beta\beta$-decay
has to be observed at different isotopes. 
Strong limits on the $0\nu\beta\beta$-decay half-life 
have been achieved in NEMO3 \cite{nemo} and CUORICINO \cite{cuore}  
experiments:
\begin{equation}
T^{0\nu}_{1/2}(^{100}\mbox{Mo}) \ge 5.8\times 10^{23}~\mbox{y},\,
T^{0\nu}_{1/2}(^{130}\mbox{Te}) \ge 3.0\times 10^{24}~\mbox{y}.
\end{equation}

After neutrino oscillations established that the neutrinos are massive, 
non-degenerate and strongly admixed, naturally most people's attention has 
focused on  the light neutrino mass mechanism of the 
$0\nu\beta\beta$-decay. 

It is well known, however, that
the $0\nu\beta\beta$-decay can be triggered by a plethora of other lepton
number violating (LNV) mechanisms. Among these we should mention the  exchange of 
 heavy neutrinos,
the exchange of SUSY superpartners with R-parity violating, 
leptoquarks, right-handed W-bosons or Kaluza-Klein excitations, 
among others, which have been discussed in the literature. 

So the day after the $0\nu\beta\beta$-decay is observed and, hopefully, established 
in a number of nuclei, the main question will be what is the dominant
mechanism that triggers the decays.

Possibilities to distinguish at least some of the possible mechanisms 
include the analysis of angular correlations between the emitted 
electrons \cite{acorr}, study of the branching ratios of 
$0\nu\beta\beta$-decays to ground and excited states \cite{ratio}, 
a comparative study of the $0\nu\beta\beta$-decay and neutrinoless 
electron capture with emission of positron ($0\nu$EC$\beta^+$) \cite{hirsch} 
and analysis of possible connections with other lepton-flavor 
violating processes (e.g., $\mu \rightarrow e \gamma$) \cite{egamma}.  

The main disadvantages of the above approaches are: small $0\nu\beta\beta$-decay 
rates to excited states,  suppressed $0\nu EC\beta^+$-decay rates, 
 experimental challenges to observe the produced X-rays 
or Auger electrons and the fact that most 
double $\beta$-decay experiments of the next generation 
are not sensitive to electron tracks.

In this paper we shall analyze what happens, if several mechanisms 
are active for the $0\nu\beta\beta$-decay. We will show that 
all LNV parameters, including the most interesting mass term, can be determined provided that  $0\nu\beta\beta$ data 
from traditional experiments involving a sufficient number of nuclear targets become available.

\section{The coexistence of few LNV mechanisms of the $0\nu\beta\beta$-decay}

The subject of interest is a coexistence  of the following 
LNV mechanisms of the $0\nu\beta\beta$-decay: 
i) Light neutrino mass mechanism. 
ii) Heavy neutrino mass mechanism. Both mechanisms assume only
left-hand current weak interactions. iii) The trilinear R-parity 
breaking SUSY mechanism generated by gluino exchange. For the sake
of simplicity we shall assume that the lepton violating parameters are relatively real as, e.g., is the the situation in the case CP conservation.
 
The inverse value of the $0\nu\beta\beta$-decay half-life for 
a given isotope $(A,Z)$ can be written as
\begin{eqnarray}
\frac{1}{T^{0\nu}_{1/2}} &=& G^{0\nu}(E_0,Z) 
|\eta_{\nu} {M}^{0\nu}_\nu + \eta_{N} {M}^{0\nu}_N + \eta_{R_p \hspace{-1em}/\;\:} 
{M}^{0\nu}_{R_p \hspace{-1em}/\;\:}|^2.\nonumber\\
\label{eq.1}
\end{eqnarray}
Here, $\eta_{\nu, N, R_p \hspace{-1em}/\;\:}$ and ${M}^{0\nu}_{\nu, N, R_p \hspace{-1em}/\;\:}$ 
are, respectively, the
LNV parameters and the nuclear matrix elements (NMEs), in the order given above. Each of the  NMEs depends, in general, quite differently on the nuclear structure of the particular isotopes 
$(A,Z)$, $(A,Z+1)$ and $(A,Z+2)$ under study. 

$G^{0\nu}(E_0,Z)$ is the known phase-space factor 
($E_0$ is the energy release), which include fourth
power of axial-coupling constant $g_A = 1.25$.
The $G^{0\nu}(E_0,Z)$ contain the inverse
square of the nuclear radius $R^{-2}$, compensated by the
factor $R$ in ${M}^{0\nu}$. The assumed value of the 
nuclear radius is $R = r_0 A^{1/3}$ with $r_0 = 1.1~fm$.
The phase-space factors are tabulated in Ref. \cite{Sim99}. 

The lepton number violating mechanisms of interest 
together with corresponding nuclear matrix elements
are presented briefly below.

\subsection{Light Majorana neutrino exchange mechanism}

In the case of light-neutrino mass mechanism of 
the $0\nu\beta\beta$-decay we have
\begin{eqnarray}
\eta_{\nu} = \frac{m_{\beta\beta}}{m_e}
\end{eqnarray}
Under the assumption of the mixing of three light massive Majorana
neutrinos the effective Majorana neutrino mass 
$\langle m_{\beta\beta} \rangle$ takes the form
\begin{equation}
m_{\beta\beta}  = \sum_i^{3} |U_{ei}|^2 \xi^{CP}_i m_i ~,
~({\rm all~} m_i \ge 0)~,
\end{equation}
where $U_{ei}$ is the first row of the neutrino mixing matrix and
$\xi^{CP}_i$ are unknown Majorana CP phases. $m_i$ is the light 
neutrino mass ($m_i \le 1$ eV, i=1, 2, 3). In this case only left-handed
weak interaction is taken into account.

The nuclear matrix element 
${M}^{0\nu}_{\nu}$ consists of Fermi, Gamow-Teller 
and tensor parts as
\begin{eqnarray}
{M}^{0\nu}_{\nu} =  
- \frac{M_{F(\nu)}}{g^2_A}
+ M_{GT(\nu)} + M_{T(\nu)}.
\end{eqnarray}
Here, $g_A$ is axial-vector coupling constant. 
The Fermi, Gamow-Teller and tensor operators are defined in the usual way (see Eq. (\ref{Eq:spinstr}) below) with 
exchange potentials as given elsewhere \cite{anatomy}

\subsection{Heavy Majorana neutrino exchange mechanisms}

We assume that the neutrino mass spectrum include heavy Majorana
states $N$ with masses $M_k$ much larger than the energy
scale of the $0\nu\beta\beta$-decay, $M_k \gg 1$ GeV. These
heavy states can mediate this process as the previous 
light neutrino exchange  mechanism. The difference is that the 
neutrino propagators in the present case can be contracted to points 
and, therefore, the corresponding effective transition operators are 
local unlike in the light neutrino exchange mechanism with long 
range internucleon interactions. 

The corresponding LNV parameter is given by
\begin{eqnarray}
\eta_{_N}
~&=& ~ \sum^{6}_{k=4}~ |U_{ek}|^2 ~ {\xi'}_k ~
\frac{m_p}{M_k}.
\end{eqnarray}
Here, $m_p$ is the mass of proton.  $U_{ek}$ are
elements of the neutrino mixing matrix associated with left-handed
current interactions. ${\xi'}_k$ are CP violating phases. 

Separating the
Fermi (F), Gamow-Teller (GT) and the tensor (T) contributions we
write down
\begin{eqnarray}
{\cal M}^{0\nu}_{_N} &=& - \frac{M_{F(N)}}{g^2_A}
+ M_{GT(N)} + M_{T(N)} \nonumber \\
&=&
\langle 0^+_i|\sum_{kl} \tau^+_k \tau^+_l
\left [ {H^{(N)}_F(r_{kl})}/{g^2_A} \right .\nonumber\\
&& \left .~~+ H^{(N)}_{GT}(r_{kl}) \sigma_{kl} 
- H^{(N)}_T(r_{kl}) S_{kl} \right ]
|0^+_f\rangle ,
\end{eqnarray}
where
\begin{equation}
S_{kl} = 3({\vec{ \sigma}}_k\cdot \hat{{\bf r}}_{kl})
       ({\vec{\sigma}}_l \cdot \hat{{\bf r}}_{kl})
      - \sigma_{kl}, ~~~ \sigma_{kl}=
{\vec{ \sigma}}_k\cdot {\vec{ \sigma}}_l.
\label{Eq:spinstr}
\end{equation}
The radial parts of the exchange potentials can be found elsewhere \cite{Sim99}.

\subsection{R-parity breaking SUSY mechanism}

In the SUSY models with R-parity non-conservation one encounters
 LNV couplings which  may also trigger the $0\nu\beta\beta$ decay.
Recall, that R-parity is a multiplicative
quantum number defined by $R=(-1)^{2S+3B+L}$ (S,B,L are spin, baryon
and lepton number). Ordinary particles have $R=+1$ while their
superpartners $R=-1$.
The LNV couplings emerge in this class of SUSY models from
the R-parity breaking part of the superpotential
\begin{equation}
W_{R_{p}\hspace{-0.8em}/\;:}=\lambda _{ijk}L_{i}L_{j}E_{k}^{c}+\lambda
_{ijk}^{\prime }L_{i}Q_{j}D_{k}^{c} + \mu_i L_i H_2,  
\label{W-Rp}
\end{equation}
where $L$, $Q$ stand for lepton and quark $SU(2)_{L}$
doublet left-handed superfields, while $E^{c},D^{c}$ for
lepton and down quark singlet superfields. This results in 
a lepton violating parameter entering the neutrinoless double beta decay, $\eta_{R_p \hspace{-1em}/\;\:}$.

For simplicity  we concentrate below  on the trilinear $\lambda'$ couplings and write 
$\eta_{R_p \hspace{-1em}/\;\:}=\eta_{\lambda'}$. Under  reasonable assumptions 
the  gluino exchange dominates \cite{wodecki}. We have
\begin{equation}
\eta_{\lambda'} =
\frac{\pi \alpha_s}{6}
\frac{\lambda^{'2}_{211}}{G_F^2 m_{\tilde d_R}^4}
\frac{m_p}{m_{\tilde g}}\left[
1 + \left(\frac{m_{\tilde d_R}}{m_{\tilde u_L}}\right)^2\right]^2.
\label{eta-N}
\end{equation}
Here, $G_F$ is the Fermi constant,
$\alpha_s = g^2_3/(4\pi )$ is $\rm SU(3)_c$ gauge coupling constant.
$m_{{\tilde u}_L}$, $m_{{\tilde d}_R}$ and $m_{\tilde g}$ are masses 
of the u-squark, d-squark and gluino, respectively.  


At the hadron level we assume dominance of the pion-exchange mode
\cite{wodecki,FKSS97}. 
We denote the $0\nu\beta\beta$-decay nuclear matrix element, 
${\cal M}^{0\nu}_{R_p \hspace{-1em}/\;\:}$,
of Eq. (\ref{eq.1}) as ${\cal M}^{0\nu}_{\lambda'}$ with
\begin{equation}
{\cal M}^{0\nu}_{\lambda'} = c_A \Big[
 \frac{4}{3}\alpha^{1\pi}\left(M_{T}^{1\pi} - M_{GT}^{1\pi} \right)
      +
      \alpha^{2\pi}\left(M_{T}^{2\pi} - M_{GT}^{2\pi} \right)\Big]\,
\end{equation}
with $c_{_{A}} = m^2_{_{A}}/(m_p m_e)$ ($m_A = 850$ MeV). The
structure coefficients of the one-pion  $\alpha^{1\pi}$ and
two-pion mode $\alpha^{2\pi}$ are \cite{wodecki,FKSS97}:
$\alpha^{1\pi} = -0.044$ and $\alpha^{2\pi} = 0.20 $. The partial
nuclear matrix elements of the $R_p \hspace{-1em}/\;\:$  SUSY
mechanism for the $0\nu\beta\beta$-decay process are:
\begin{eqnarray}
{M}_{GT}^{k\pi} &=&
\langle 0^+_f|~\sum_{k\neq l} ~\tau_k^+ \tau_l^+ ~
~H_{GT}^{k\pi}(r_{kl})
~{\bf{\sigma}}_i\cdot{\bf{\sigma}}_j,
~| 0^+_i \rangle , \nonumber \\
{M}_{T}^{k\pi} &=&
\langle 0^+_f|~\sum_{k\neq l} ~\tau_k^+ \tau_l^+ ~
~H_{T}^{k\pi}(r_{kl})
~{S}_{kl}
~| 0^+_i \rangle 
\end{eqnarray}
with the radial functions given elsewhere \cite{wodecki,FKSS97}.  
Under these assumptions  the obtained nuclear matrix elements are given in table \ref{table.1}

\begin{table}[t]
  \begin{center}
\caption{\label{table.1} 
The $0\nu\beta\beta$-decay NMEs calculated with the Selfconsistent
Renormalized Quasiparticle Random Phase Approximation (SRQRPA). 
The Coupled Cluster Method (CCM)
short-range correlations calculated with CD-Bonn  potential are taken
into account \protect\cite{src}. $g_A = 1.25$.}
\begin{tabular}{lcccc}
\hline\hline
Nucl. trans. &  $G^{0\nu}(E_0,Z)~[y^{-1}]$ & 
$M^{0\nu}_\nu$ & $M^{0\nu}_N$  &  $M^{0\nu}_{\lambda'}$ \\\hline 
$^{76}$Ge$\rightarrow ^{76}$Se  &
  $7.98\times 10^{-15}$ & 5.82  & 412. & 596. \\
$^{100}$Mo$\rightarrow ^{100}$Ru &
  $5.73\times 10^{-14}$ & 5.15  & 404. & 589. \\
$^{130}$Te$\rightarrow ^{130}$Xe &
  $5.54\times 10^{-14}$ & 4.70  & 385. & 540. \\ 
\hline\hline
\end{tabular}
  \end{center}
\end{table}


In obtaining the nuclear matrix elements we used 
the Self-consistent Renormalized Quasiparticle Random Phase
Approximation  (SRQRPA) \cite{srpa} to calculate nuclear matrix elements
(NMEs) $M^{0\nu}_\nu$, $M^{0\nu}_N$ and $M^{0\nu}_{\lambda'}$. 
The SRQRPA takes into account the Pauli exclusion principle and 
conserves the mean particle number in correlated ground state.
For A=76 and 100 nuclear systems the  single-particle model space
consists of $0-5\hbar\omega$ oscillator shells and for A=130 
nuclear system we used $0-5\hbar\omega$ shells  plus $0i_{11/2}$ 
and $0i_{13/2}$ levels both for protons and  neutrons. 
In the calculation of the $0\nu\beta\beta$-decay NMEs  
the two-nucleon short-range correlations  
derived from same potential as residual interactions,
namely from the CD-Bonn potential \cite{src}, are considered.
The calculated NMEs are given in Table \ref{table.1}.

%

\section{Calculation and discussion}

\subsection{Dominance of a single $0\nu\beta\beta$-decay mechanism}

Commonly, it is assumed that a single LNV mechanism is responsible
for the $0\nu\beta\beta$-decay. Let suppose it is the light neutrino
mass ($m_{\beta\beta}$) or heavy neutrino mass ($\eta_N$) mechanism. 
Then, the $0\nu\beta\beta$-decay half-lives $T_1$ and $T_2$ for two 
nuclear systems are related with
equation
\begin{equation}
|m_{\beta\beta}| = \frac{m_e}{|M^\nu_i| \sqrt{T_i~ G_i}}, \,
|\eta_N| = \frac{1}{|M^{\eta_N}_i| \sqrt{T_i~ G_i}}
\label{one.masseta}
\end{equation}
Here, $G_{i}$ is the kinematical factor, while $M^\nu_{i}$ and $M^{\eta_N}_{i}$ are
nuclear matrix elements associated with $m_{\beta\beta}$  and $\eta_N$ parameters,
respectively for the target $i$. 

\subsection{Two active $0\nu\beta\beta$-decay mechanisms}

\begin{figure}[htb!]
\includegraphics[width=.42\textwidth,angle=0]{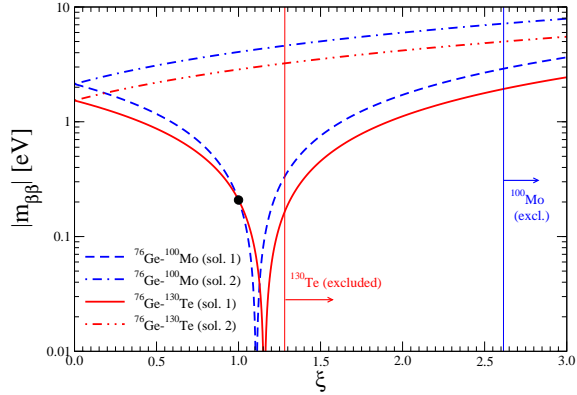}
\caption{The effective Majorana mass of neutrinos in the case
of two active mechanisms of the $0\nu\beta\beta$-decay, 
namely light and heavy neutrino exchange mechanisms, as function
of parameter $\xi$ (see Eq. (\ref{xi})). 
$T^{0\nu}_{1/2}(^{76}Ge) = 2.23\times 10^{25}~$y \protect\cite{evidence}
is assumed. Solutions 1 and 2 were obtained for equal and opposite 
signs on the left hand side of Eqs. (\ref{2eqs}), respectively.
The bold point indicates the value of $m_{\beta\beta}$, if the
light neutrino exchange is the only active mechanism.
} 
\label{fig:1}
\end{figure}

We will now move into the case of two competing $0\nu\beta\beta$-decay mechanisms
representing by the LNV parameters $m_{\beta\beta}$ and $\eta$, $\eta$ could be $\eta_N$ or $\eta_{R_p \hspace{-1em}/\;\:}$. In this case
we have four different sets of two linear equations: 
\begin{equation}
 \frac{\pm1}{\sqrt{T_1~ G_1}} = 
\frac{m_{\beta\beta}}{m_e} M^\nu_1 + \eta  M^{\eta}_1, 
 \frac{\pm1}{\sqrt{T_2~ G_2}} = \frac{m_{\beta\beta}}{m_e} 
M^\nu_2 + \eta M^{\eta}_2.
\label{2eqs}
\end{equation}
For the absolute value of the LNV parameters we find two different solutions,
\begin{eqnarray}
|m_{\beta\beta}| &=&~ \left| \frac{m_e}{M^\nu_1 \sqrt{T_1~ G_1}}
\frac{M^\nu_1~ M^{\eta}_2}{(M^\nu_1 M^{\eta}_2 - M^\nu_2 M^{\eta}_1)}\right.\nonumber\\
        && \left. \pm \frac{m_e}{M^\nu_2 \sqrt{T_2~ G_2}} 
\frac{M^\nu_2~ M^{\eta}_1}{(M^\nu_1 M^{\eta}_2 - M^\nu_2 M^{\eta}_1)}\right|
\label{mbb2e}\\
|\eta| &=&~   \left|\frac{1}{M^{\eta}_1 \sqrt{T_1~ G_1}}
\frac{M^\eta_1~ M^\nu_2}{(M^{\eta}_1 M^\nu_2 - M^{\eta}_2 M^\nu_1)}\right.\nonumber\\
        && \left. \pm \frac{1}{M^{\eta}_2 \sqrt{T_2~ G_2}} 
\frac{M^\eta_2~ M^\nu_1}{(M^{\eta}_1 M^\nu_2 - M^{\eta}_2 M^\nu_1)}\right|,
\label{eta2e}
\end{eqnarray}
 We note, however,  that for $\eta = 0$ 
Eqs. (\ref{mbb2e}) and (\ref{eta2e}) are reduced to Eq. (\ref{one.masseta}).

By assuming now $\eta \equiv \eta_N$ the solutions for $|m_{\beta\beta}|$ will be analyzed for two different
combinations of nuclear systems, namely with A = 76 and 100 (case I) and 
A = 76 and 130 (case II).
An additional assumption is that the $0\nu\beta\beta$-decay  
half-life of $^{76}$Ge has been measured with 
$T^{0\nu}_{1/2}(^{76}$Ge) given in (\ref{evid}).
In Fig. \ref{fig:1} the two solutions for $|m_{\beta\beta}|$
are plotted as function of $\xi$, where
\begin{equation}
\xi = \frac{|M^\nu_1| \sqrt{T_1~ G_1}}{|M^\nu_2| \sqrt{T_2~ G_2}}, 
\label{xi}
\end{equation}
Indices 1 and 2 denote the above cases I and II 
respectively.
The parameter $\xi$ represents the unknown half-life of the 
$0\nu\beta\beta$-decay of $^{100}$Mo or $^{130}$Te. We note that for 
$\xi = 1$ the solution for active only light neutrino mass mechanism 
given by Eq. (\ref{one.masseta}) is reproduced and that $\xi = 0$
means non-observation of the $0\nu\beta\beta$-decay for a considered
isotope. 

By glancing the Fig. \ref{fig:1} we see that for both combinations of nuclear 
systems the two solutions for $|m_{\beta\beta}|$ exhibit  similar behavior. 
For $\xi\approx 1.1$ one of the solution can be even equal to zero.
The second solution predicts $m_{\beta\beta} > 1~eV$. The current 
lower limits on the half-life of the  $0\nu\beta\beta$-decay of $^{100}$Mo 
or $^{130}$Te restrict the effective mass of Majorana neutrinos to intervals 
$(0.0~$eV, $7.2~$eV) and $(0.0~$eV, $3.2~$eV), respectively. So, within the 
considered assumptions the claim of evidence of the $0\nu\beta\beta$-decay 
of $^{76}$Ge \cite{evidence}
is compatible also with inverted ($m_i < 50~$meV)
or normal ($m_i \approx$ few~meV) hierarchy of neutrino
masses. From Fig. \ref{fig:1} it follows that a small improvement of the
current half-life limit for $^{130}$Te up to value 
$4.1\times 10^{24}$ y ($\xi_{\text{Te}}\simeq 1.1$) would exclude 
these possibilities. Another finding is that
the non-observation of the $0\nu\beta\beta$-decay for $^{100}$Mo or $^{130}$Te 
(i.e., $\xi = 0$)  cannot rule out the claim for evidence of 
the $0\nu\beta\beta$-decay of $^{76}$Ge. This can only happen if,
in a more sensitive Ge  experiment like GERDA or Majorana, no
$0\nu\beta\beta$-decay signal will be registered. 

\vspace{0.8cm}
\begin{figure}[t]
\includegraphics[height=250pt,width=.47\textwidth,angle=0]{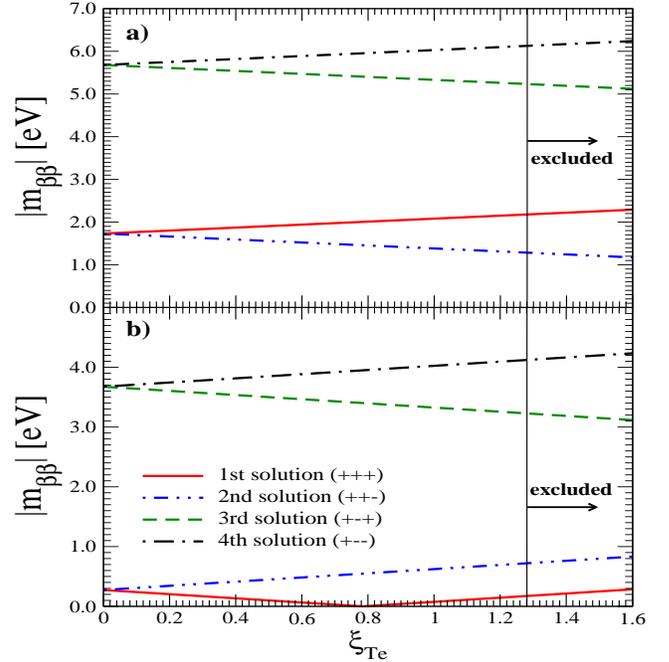}
\caption{The effective Majorana mass of neutrinos in the case
of three active mechanisms of the $0\nu\beta\beta$-decay, 
namely light and heavy neutrino exchange mechanisms and R-parity 
breaking SUSY mechanism with gluino exchange, as function
of parameter $\xi_{\text{Te}}$. Nuclear systems with A = 76, 100 and 130 
are considered.
$T^{0\nu}_{1/2}(^{76}$Ge) is the same as in Fig. \ref{fig:1}.
$\xi_{\text{Mo}} = 2.6$ and $\xi_{\text{Mo}} = 1.2$ are assumed in upper (a) 
and lower (b) panels, respectively. For each solution there are
given in brackets signs in front of term $1/\sqrt{T G}$ 
in Eqs.(\ref{3lineq}) for ${^{76}Ge}$, ${^{100}Mo}$  and
${^{130}Te}$.} 
\label{fig:2}
\end{figure}

\subsection{Three active  $0\nu\beta\beta$-decay mechanisms}

In the case of three active $0\nu\beta\beta$-decay mechanisms
represented by the LNV parameters $m_{\beta\beta}$, $\eta_N$
and $\eta_{\lambda'}$ assuming the  measurement of the life time of the 
$0\nu\beta\beta$-decay of three isotopes one obtains a  set of
three linear equations: 
\begin{equation}
\frac{\pm 1}{\sqrt{T_i~ G_i}} = 
\frac{m_{\beta\beta}}{m_e} M^\nu_i + \eta_N M^{\eta}_i + \eta_{\lambda'} M^{\lambda'}_i,\, i=1,2,3 
\label{3lineq}
\end{equation}

The equations (\ref{3lineq}) admit a set of four different solutions, which are exhibited  in Fig. \ref{fig:2}. The upper and lower panels correspond 
to $\xi_{\text{Mo}} = 2.6$ (current limit) and   $\xi_{\text{Mo}} = 1.2$, respectively. 
The allowed ranges  of $m_{\beta\beta}$ are as follows:
i) $(1.3, 2.2)$ eV and  $(5.2, 6.1)$ eV for upper panel,
ii) $(0.0, 1.7)$ eV and  $(3.2, 4.1)$ eV for lower panel.
We see that for a given value of $\xi_{\text{Te}}$ 
allowed intervals for $|m_{\beta\beta}|$ depend strongly  on the value 
of $\xi_{\text{Mo}}$. The upper two solutions determining the second interval
are already excluded by the Mainz and Troitsk tritium experiments 
\cite{otten}. In case the claim of evidence of the $0\nu\beta\beta$
decay of $^{76}$Ge would be ruled out by other experiments, i.e. for larger value of 
$T^{0\nu}_{1/2}(^{76}$Ge), they would decrease and might be important.

\section{Conclusions}

It has been shown that the extraction of the most important neutrino mass  contribution, 
entering neutrinoless double beta decay, can be disentangled from the other mechanisms, 
if and when the decay rates in a sufficient number of nuclear targets become available. 
In the present calculation, to simplify the exposition, we restricted ourselves in 
the special case of no right handed currents and made the assumption that the LNV 
parameters are relatively real. To be more specific, in addition to the standard 
light neutrino mass mechanism of the $0\nu\beta\beta$-decay, we considered two 
additional  LNV mechanisms, namely those involving the  exchange of heavy neutrinos 
and R-parity breaking SUSY  with gluino exchange. We find that this improved analysis 
leads to completely different results compared to those of one mechanism at a time. 
It is now possible that larger values of $|m_{\beta\beta}|$ can be consistent with the 
data, since the contribution of the other mechanisms could be interfering with it destructively.

We specifically discussed the extracted value of the effective Majorana neutrino mass , $m_{\beta\beta}$,  
assuming the claim of evidence
of the $0\nu\beta\beta$-decay of $^{76}$Ge \cite{evidence}  as a function of
 half-life data for the two promising nuclei, ($^{100}$Mo and
$^{130}$Te). We showed that in an analysis including
two and three nuclear systems there are 2 and 4 different possible solutions
for $|m_{\beta\beta}|$, respectively.
One of the solutions leads to small values of $|m_{\beta\beta}|$, 
when all mechanisms add up coherently. This is compatible also with inverted ($m_i < 50~$meV)
or normal ($m_i \approx$ few~meV) hierarchy of neutrino
masses.
Other solutions, however, allow quite large values of 
$|m_{\beta\beta}|$, even larger than 1 eV. These can, of course, be excluded by cosmology
and tritium $\beta$-decay experiments. It may not, however, be possible to exclude 
these  solutions, if the claim of evidence for $^{76}$Ge would be ruled out by future experiments, 
since, then, the values we obtain become smaller than those of the other experiments.

It is thus important that experiments involving as many different targets as possible 
be pursued. Furthermore, in the presence of interference between the various mechanisms, 
the availability of reliable  nuclear matrix elements becomes more imperative.

\section{Acknowledgments}

This work was supported in part by the DFG project 436 SLK 17/298, 
the Transregio Project TR27 "Neutrinos and Beyond", 
by the Graduiertenkolleg GRK683,  by the VEGA 
Grant agency  under the contract No.~1/0249/03 and 
by a Humboldt Research Award. Partial support was 
also provided by the EU network MRTN-CT-2006-035683 (UniverseNet).

\end{document}